\documentclass[journal]{IEEEtran}

\usepackage{booktabs} 
\usepackage{multirow}
\usepackage{todonotes}
\usepackage{verbatim}

\usepackage{bm}
\usepackage{amsmath}
\usepackage{amssymb}
\usepackage{amsfonts}
\usepackage{mathtools}
\usepackage{stmaryrd}
\usepackage{graphicx}
\usepackage{color}
\usepackage{colortbl}

\usepackage{comment}
\usepackage{proof}
\usepackage{fancybox}

\usepackage{wrapfig}
\renewcommand{\phi}{\varphi}

\newcommand{\R}{\mathbb{R}}

\newcommand{\Rpos}{\R_{+}}

\newcommand{\Rm}{\R^{m}}
\newcommand{\Rin}{\R^{m}}
\newcommand{\Rout}{\R^{n}}

\newcommand{\argmax}{\mathop{\rm arg~max}\limits}

\newcommand{\argminHillClimb}{\mathop{\rm arg~min^{\rm HillClimb}}\limits}

\newcommand{\bv}{\mathbf{v}}
\newcommand{\bw}{\mathbf{w}}
\newcommand{\bu}{\mathbf{u}}
\newcommand{\sem}[1]{\llbracket #1 \rrbracket}

\newcommand{\DiaOp}[1]{\Diamond_{#1}}
\newcommand{\BoxOp}[1]{\square_{#1}}
\newcommand{\speed}{\mathit{speed}}
\newcommand{\gear}{\mathit{gear}}
\newcommand{\rpm}{\mathit{rpm}}
\newcommand{\AF}{\mathit{AF}}
\newcommand{\AFref}{\mathit{AFref}}

\newcommand{\STL}{\textrm{STL}}
\newcommand{\myparagraph}[1]{\paragraph*{#1}}
\newcommand{\Var}{\mathbf{Var}}

\newcommand{\UntilOp}[1]{\mathbin{\mathcal{U}_{#1}}}

\newcommand{\Rnn}{\R_{\ge 0}}

\newcommand{\Defeq}{:=}
\newcommand{\Robust}[2]{{ \llbracket #1, #2 \rrbracket}}
\newcommand{\Vee}[1]{{{\bigsqcup_{#1}}}}
\newcommand{\Wedge}[1]{{{\bigsqcap_{#1}}}}
\newcommand{\throttle}{\mathit{throttle}}
\newcommand{\brake}{\mathit{brake}}

\newcommand{\tbcolor}{\cellcolor{green!25}}
\newcommand{\tbgray}{\cellcolor{gray!25}}

\usepackage{algorithm}
\usepackage[noend]{algpseudocode} 
\usepackage{etoolbox}

\usepackage{gnuplot-lua-tikz}
\usepackage{pgfplotstable} 
\pgfplotsset{compat=1.12}
\usepackage{booktabs}
\usepackage{tabularx}

\usepackage{arydshln}
\usepackage{hyperref}

\usepackage{enumitem}
\setlistdepth{20}
\renewlist{itemize}{itemize}{20}
\setlist[itemize]{label=\textbullet}

\usepackage[all,pdftex]{xy}
\CompileMatrices 
\xyoption{v2}
\xyoption{curve}
\xyoption{2cell}
\SelectTips{cm}{}  
\UseAllTwocells

\newtheorem{mytheorem}{Theorem}[section]
\newtheorem{mydefinition}[mytheorem]{Definition}

\begin{document}
\title{Two-Layered Falsification of Hybrid Systems Guided by Monte Carlo Tree Search}

\author{Zhenya~Zhang,
        Gidon~Ernst,
        Sean~Sedwards,
        Paolo~Arcaini,
        and~Ichiro~Hasuo
\thanks{Zhenya Zhang, Paolo Arcaini and Ichiro Hasuo are with the National Institute of Informatics, Japan. Zhenya Zhang and Ichiro Hasuo are also with the Graduate University for Advanced Studies (SOKENDAI), Japan.}
\thanks{Gidon Ernst was at the National Institute of Informatics, Japan during this work; he is now with the University of Melbourne, Australia.}
\thanks{Sean Sedwards is with the University of Waterloo, Canada.}
\thanks{This article was presented in the International Conference on Embedded Software 2018 and appears as part of the ESWEEK-TCAD special issue.}
\thanks{Manuscript received ; revised .}}

\markboth{Journal of \LaTeX\ Class Files,~Vol.~14, No.~8, August~2015}
{Shell \MakeLowercase{\textit{et al.}}: Bare Demo of IEEEtran.cls for IEEE Journals}

\maketitle

\begin{abstract}
Few real-world hybrid systems are amenable to formal verification, due to
 their complexity and black box components. \emph{Optimization-based
 falsification}---a  methodology of search-based testing that employs
 stochastic optimization---is thus attracting attention as an alternative
 quality assurance method. Inspired by the recent work that advocates
 \emph{coverage} and \emph{exploration} in falsification, we introduce a
 two-layered optimization framework that uses \emph{Monte Carlo tree
 search (MCTS)}, a popular machine learning technique with
 solid mathematical and empirical foundations (e.g.\ in computer Go). MCTS is used in the upper layer of our framework; it
 guides the lower layer of local hill-climbing optimization, thus
 balancing exploration and exploitation in a disciplined manner. We demonstrate the proposed framework through experiments with benchmarks from the automotive domain. 
\end{abstract}


\IEEEpeerreviewmaketitle

\section{Introduction}\label{sec:introduction}
\subsection{Hybrid Systems} 
\IEEEPARstart{Q}{uality} assurance  of \emph{cyber-physical systems} (CPS) is a problem of great interest. Errors in CPS, such as cars and aircrafts, can lead to economic and social damage, including loss of human lives. Unique challenges in quality assurance are posed by the nature of CPS: in the form of
 {\em hybrid systems} they 
comprise the discrete dynamics of computers and the continuous dynamics of physical components. Continuous dynamics combined with other features, such as complexity (a modern car can contain $10^{8}$ lines of code) and black-box components (such as parts coming from external suppliers), make it very hard to apply formal verification to CPS.

\begin{figure}
   \centering
     \includegraphics[width=.45\textwidth]{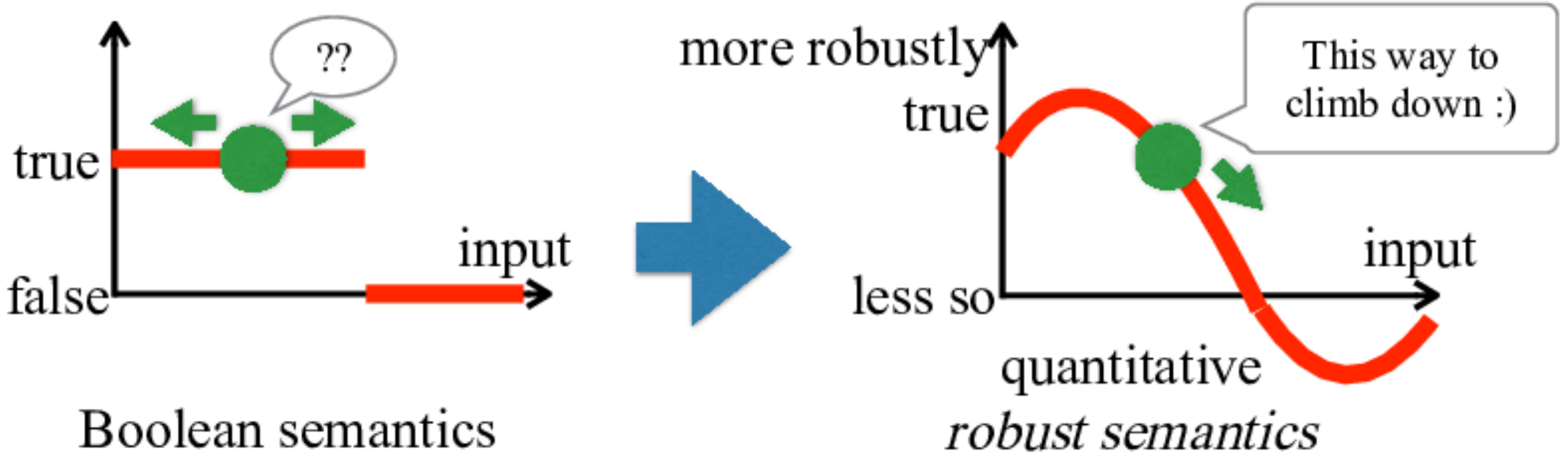}
    \caption{From Boolean to robust semantics}
    \label{fig:fromQualitativeToQuantitative}
\end{figure}

An increasing number of  researchers and practitioners are therefore turning to  \emph{optimization-based falsification} as a quality assurance measure for CPS.
The problem is formalized as follows. 
\begin{center}
   \begin{minipage}{0.45\textwidth}
    \underline{\bfseries The falsification problem}
    \begin{itemize}
    \item{\textbf{Given:}} 
      a \emph{model} $\mathcal{M}$ (that takes an input signal $\bu$
      and  yields an output signal $\mathcal{M}(\bu)$), and
      a \emph{specification} $\varphi$ (a temporal formula)
    \item{\textbf{Find:}} 
      an \emph{error input}, that is, an input signal $\bu$ such
      that the corresponding output $\mathcal{M}(\bu)$ violates $\varphi$ 
    \end{itemize}
  \end{minipage}
\begin{math}
   \xymatrix@1@+1.2em{
   {}
     \ar[r]^-{\bu}
   &
   {\hspace{+0.3em}\xybox{ *++[F]{\mathcal{M}} }}
     \ar[r]^-{\mathcal{M}(\bu)}_-{\not\models\varphi \; ?}
   &
   {}
   }
  \end{math}
\end{center}
In the optimization-based falsification approach, the above falsification problem is turned into an optimization problem. This is possible thanks to \emph{robust semantics} of temporal formulas~\cite{FainekosP09}. Instead of the Boolean satisfaction relation $\bv\models\varphi$, robust semantics assigns a quantity $\sem{\bv,\varphi}\in\R\cup\{\infty,-\infty\}$ that tells us, not only whether $\varphi$ is true or not (by the sign), but also \emph{how robustly} the formula is true or false. This allows one to employ hill-climbing optimization (see Fig.~\ref{fig:fromQualitativeToQuantitative}): we iteratively generate input signals, in the direction of decreasing robustness, hoping that eventually we hit negative robustness. 

Optimization-based falsification is a subclass of \emph{search-based testing}: it adaptively chooses test cases (input signals $\bu$) based on previous observations. One can use stochastic algorithms for optimization, such as
 simulated annealing (SA), globalized Nelder-Mead (GNM~\cite{LuersonLeRiche2004}) 
and covariance matrix adaptation evolution strategy (CMA-ES~\cite{AugerH05}), which turn out to be much more
 scalable than  model checking algorithms that rely on exhaustive
 search. Note also that the system model $\mathcal{M}$ can be  black
 box:  observing the correspondence between input $\bu$ and output
 $\mathcal{M}(\bu)$ is enough.
 Observing an error $\mathcal{M}(\bu')$ for some input $\bu'$ is sufficient evidence for a system designer  to know that the system needs improvement. 
Besides these practical advantages, optimization-based falsification is an interesting topic from a scientific point of view,  combining formal and structural reasoning with stochastic optimization.

The approach of  optimization-based falsification was initiated in~\cite{FainekosP09} and has been actively pursued ever since~\cite{Annpureddy-et-al2011,AdimoolamDDKJ17,DeshmukhJKM15,KuratkoR14,Donze10,DonzeM10,DreossiDDKJD15,ZutshiDSK14,AkazakiKH17,SilvettiPB17,DreossiDS17}. See~\cite{KapinskiDJIB16} for a recent survey. There are now mature tools, such as Breach~\cite{Donze10} and S-Taliro~\cite{Annpureddy-et-al2011}, which work with industry-standard Simulink models.

\subsection{The Exploration-Exploitation Trade-off in  Falsification}
In op\-ti\-mi\-za\-tion-based falsification, the important role of \emph{coverage} is advocated by many authors~\cite{KuratkoR14,DeshmukhJKM15,AdimoolamDDKJ17,DreossiDDKJD15} (see also~\S{}\ref{sec:relatedwork}). One reason is that in highly nonconvex optimization problems for falsification, eager hill climbing can easily be trapped in local minima and thus fail to find an error input (i.e.\ a global minimum) that exists elsewhere. Another reason is that coverage gives a certain degree of confidence for absence of error input, in case search for error input is unsuccessful. 

This puts us in the  \emph{exploration-exploitation trade-off}, a typical dilemma in stochastic optimization and machine learning (specifically in reinforcement/active learning). While exploitation guides us to  pursue the direction that seems promising, based on the previous observations, we have to occasionally explore in order to avoid getting stuck in local minima.
Many common stochastic hill-climbing algorithms, such as
SA, GNM and CMA-ES, contain implicit exploration mechanisms. At the same time, explicit methods for exploration in falsification have been pursued e.g.\ in~\cite{KuratkoR14,DeshmukhJKM15,AdimoolamDDKJ17,DreossiDDKJD15} (see~\S{}\ref{sec:relatedwork}). 

\myparagraph{Contribution}
Our  main contribution  is, in the context of hybrid system falsification, to balance exploration and exploitation in a systematic and mathematically disciplined way using \emph{Monte Carlo tree search} (MCTS). We integrate hill-climbing optimization in MCTS, and obtain a two-layered optimization framework.

MCTS is an \emph{expected outcome}~\cite{Abramson1990} algorithm that searches a tree whose nodes are usually organized according to causal relationships, interleaving \emph{search} (walking down the already expanded tree in a promising direction) with \emph{playout} (expanding a new node and estimating its reward). 
One reason for the success of MCTS is that its search strategies nicely balance exploration and exploitation.
The most common search strategy, UCT (UCB applied to trees~\cite{KocsisS06}), is derived from the solid theoretical background of the UCB (upper confidence bounds) strategy for multi-armed bandit problems~\cite{Auer2002}.
Typical applications allowing such a structured search space are \emph{decision problems}, such as games. In particular, MCTS is attracting a lot of attention thanks to its success in computer Go~\cite{Silver-et-al2015}.
While MCTS is a relatively new methodology, it has already established its position in the rapidly growing community of machine learning. See~\cite{BrownePWLCRTPSC12} for a survey.

\begin{figure}[tbp]
\centering
      \includegraphics[width=.4\textwidth]{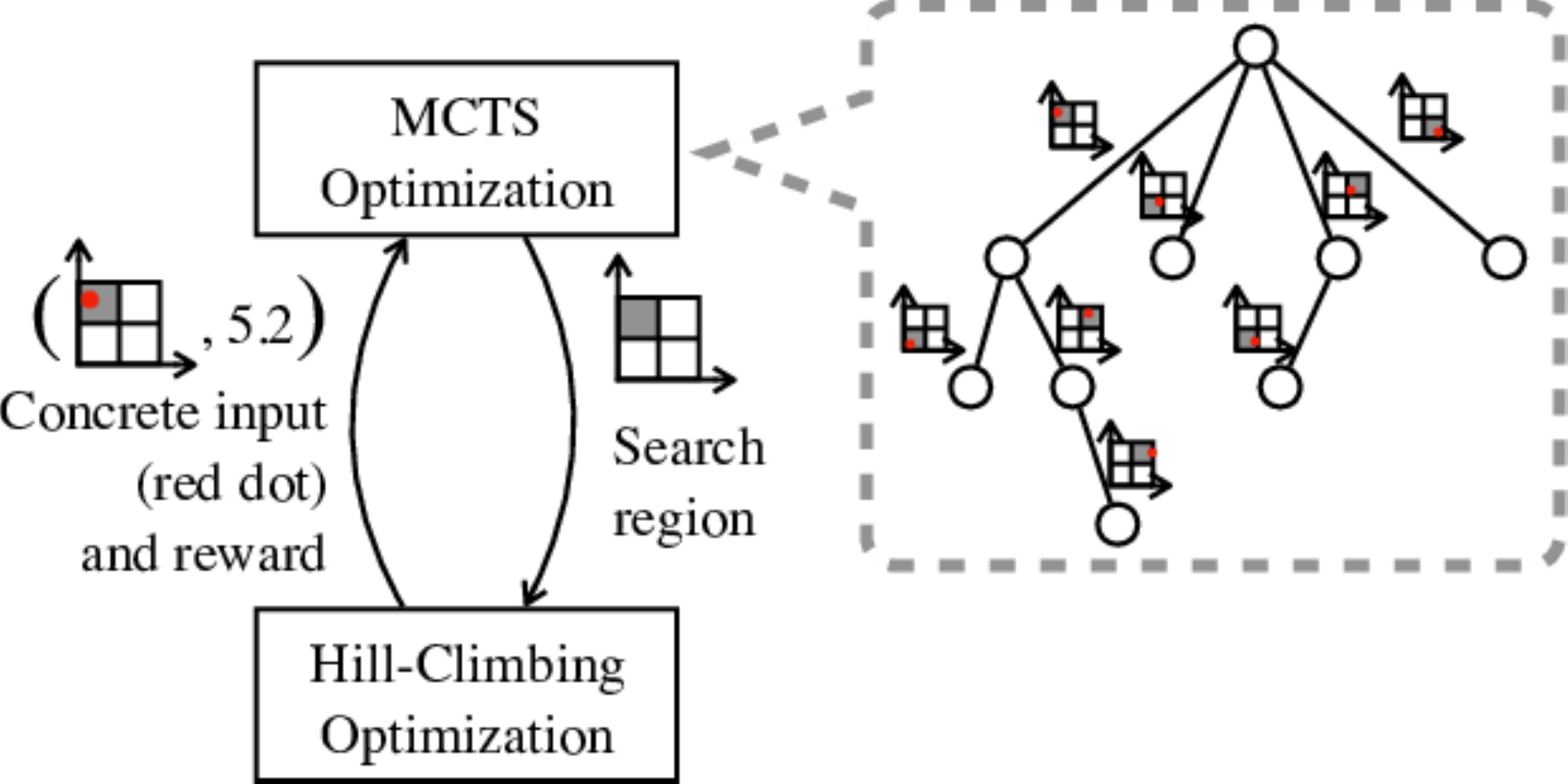}
    \caption{Our two-layered optimization framework}
    \label{fig:twoLayerd}
\end{figure}
\begin{figure}[tbp]
\centering
      \includegraphics[width=0.45\textwidth]{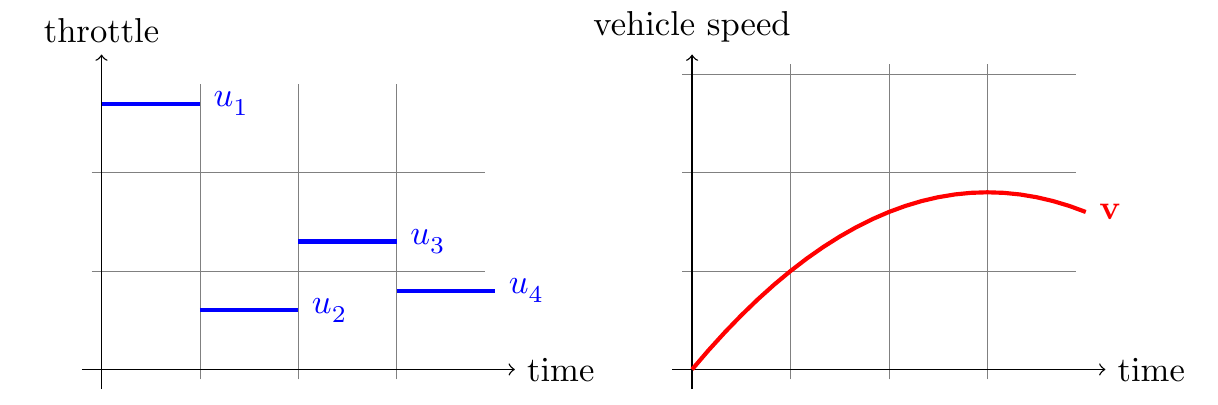}
    \caption{A piecewise constant input signal (one-dimensional, throttle, left) for a simple automotive powertrain model, and the corresponding output signal (one-dimensional, vehicle speed, right).}
   \label{fig:timeStaging}
\end{figure}

Our framework uses robustness values as rewards in MCTS, and employs hill-climbing optimization for  playout in MCTS. This way we integrate hill-climbing in Monte Carlo tree search in a systematic way. In our two-layered framework (Fig.~\ref{fig:twoLayerd}), the upper optimization layer picks (by MCTS) a region in the input space, from which a concrete input value should be sampled.
The lower layer then picks (by hill-climbing) an optimal concrete input value within the prescribed region. We also compute the robustness of the specification under the chosen  input. This value is fed back to the upper layer as a reward, which is then used by the tree search strategy to balance exploration and exploitation.

In our two-layered framework, hill-climbing optimization---whose potential in  falsification of hybrid systems has been established, see e.g.~\cite{KapinskiDJIB16}---is supervised by MCTS, with MCTS dictating which region to sample from. By expanding new children, MCTS  can tell the hill-climbing optimization to try an input region that has not yet been explored, or to  exploit  and dig deep in a direction that seems promising.
This combination of MCTS and application-specific lower-layer optimization seems to be a useful approach that can apply to problems other than hybrid system falsification. See~\S{}\ref{sec:relatedwork} for further discussion.

Our use of MCTS depends on our \emph{time-staged} approach to falsification \cite{ErnstHZS18}, in which we synthesize $K$ input segments one after another.
Those input segments are for the time intervals $[0,\frac{T}{K}), [\frac{T}{K},\frac{2T}{K}),\dotsc, [\frac{(K-1)T}{K},T]$, where $T$ is the time horizon.
The search tree will then be of depth $K$. See Fig.~\ref{fig:timeStaging}. In this paper we restrict input signals to piecewise-constant ones (this is a common assumption in falsification); an edge in the MCTS search tree from depth $i-1$ to $i$ (see Fig.~\ref{fig:twoLayerd}) determines the input value $u_{i}$ for the interval  $[\frac{(i-1)T}{K},\frac{iT}{K})$.

We have implemented our two-layered falsification framework in MATLAB, building on Breach~\cite{Donze10}.\footnote{Code obtained at \url{https://github.com/decyphir/breach}.} Our experiments with benchmarks from~\cite{HoxhaAF14,JinDKUB14,DeshmukhHJMP17} demonstrate the possible performance improvements, especially in the ability of finding rare counterexamples.

\myparagraph{Organization} In~\S{}\ref{sec:falsificationProblem} we formulate the falsification problem. In~\S{}\ref{sec:twoLayered} we present our main contribution, namely a two-layered optimization framework for falsification that combines MCTS and hill-climbing. Our experimental results are in~\S{}\ref{sec:experiments}. 
 In~\S{}\ref{sec:relatedwork} we discuss related work, locating the current work in the context of falsification and also of other applications of MCTS and related machine learning methods.
In~\S{}\ref{sec:concl} we conclude with some  directions of future research.

\myparagraph{Notations} The set of (positive, nonnegative) real numbers is denoted by $\R$ (and $\Rpos, \Rnn$, respectively). Closed and open intervals are denoted such as $[0,2]$ and $(2,3)$; $[0,2)=\{x\in \R\mid 0\le x < 2\}$ is a half-closed half-open interval. For a set $X$, $|X|$ denotes its cardinality. 

\section{Problem: Hybrid System Falsification}
\label{sec:falsificationProblem}
We formulate the problem of hybrid system falsification. We also introduce robust semantics of temporal logics~\cite{FainekosP09,DonzeM10} that allows us to reduce falsification to an optimization problem. 

 \begin{mydefinition}[time-bounded signal]\label{def:operationsOnSignals}
 Let $T\in \Rpos$ be a positive real. 
 An \emph{$m$-dimensional signal} with a time horizon $T$ is a function $\bw\colon [0,T]\to\Rm$. 

 Let $\bw\colon [0,T]\to \Rm$ and $\bw'\colon [0,T']\to\Rm$ be $m$-dimensional signals. Their \emph{concatenation} $\bw\cdot\bw'\colon [0,T+T']\to \Rm$ is an $m$-dimensional signal defined by
 \begin{math}
  (\bw\cdot\bw')(t):=
 \bw(t)
 \end{math}
 if $t\in [0,T]$, and
   $\bw'(t-T)$ if $t\in(T,T+T']$.

 Let $T_{1},T_{2}\in (0,T]$ such that $T_{1}<T_{2}$. The \emph{restriction} 
 $\bw|_{[T_{1},T_{2}]}\colon [0,T_{2}-T_{1}]\to \Rm$ of  $\bw\colon [0,T]\to \Rm$ to the interval  $[T_{1},T_{2}]$ is defined by $(\bw|_{[T_{1},T_{2}]})(t):=\bw(T_{1}+t)$. 
 \end{mydefinition}
 
 \begin{mydefinition}[system model $\mathcal{M}$]\label{def:systemModel}
 A \emph{system model}, with $m$-di\-men\-sion\-al input and $n$-di\-men\-sion\-al output,  is a function $\mathcal{M}$ that takes 
an input signal $\bu\colon [0,T]\to \Rin$ and returns a signal $\mathcal{M}(\bu)\colon [0,T]\to \Rout$. Here the common time horizon $T\in \Rpos$ is arbitrary. 
\end{mydefinition}

Some recent works, including~\cite{HoxhaDF18}, use sequences of time-stamped values as basic objects in their problem formulation, in place of continuous-time signals (as we do in the above). This difference is mostly presentational and not essential.

As a specification language we use  \emph{signal temporal logic} (\STL)~\cite{MalerN04}. We do so  for simplicity of presentation; we can also use more expressive logics such as the one in~\cite{AkazakiH15}.

In what follows $\Var$ is the set of variables. Variables stand for physical quantities, control modes, etc.  $\equiv$ denotes syntactic equality. 
\begin{mydefinition}[syntax]\label{def:syntax}
  In $\STL$, 
 \emph{atomic propositions} and 
 \emph{formulas} are defined as follows, respectively:
\begin{math}
       \alpha 
 \,::\equiv\,
        \bigl(\,f(x_1, \dots, x_n) > 0\,\bigr)
\end{math}, and 
\begin{math}
\varphi  \,::\equiv\,
\alpha \mid \bot
\mid \neg \varphi 
\mid \varphi \wedge \varphi 
\mid \varphi \UntilOp{I} \varphi
\end{math}. Here
 $f$ is an $n$-ary function $f:\R^n \to \R$, $x_1, \dots, x_n \in \Var$,
and $I$ is a closed non-singular interval in $\Rnn$, i.e.\ $I=[a,b]$ or $[a, \infty)$ where $a,b \in \R$ and  $a<b$.
\end{mydefinition}

 We omit subscripts $I$ for temporal operators if $I = [0, \infty)$. Other common connectives and operators, like $\lor,\rightarrow,\top$, $\Box_{I}$ (always) and $\Diamond_{I}$ (eventually), are introduced as abbreviations: $\Diamond_{I}\varphi\equiv\top\UntilOp{I}\varphi$ and 
$\Box_{I}\varphi\equiv\lnot\Diamond_{I}\lnot\varphi$.  Atomic formulas like $f(\vec{x})\le c$, where $c\in\R$ is a constant, are also accommodated by using negation and the function $f'(\vec{x}):=f(\vec{x})-c$.

\begin{mydefinition}[robust semantics~\cite{DonzeM10}]\label{def:semantics} For an  $n$-dimensional signal $\bw \colon \Rnn\to \R^{n}$ and $t\in \Rnn$,
 $\bw^t$ denotes the \emph{$t$-shift} of $\bw$, that is, 
 $\bw^t(t') \Defeq \bw(t+t')$.

Let $\bw \colon \Rnn \to \R^{|\Var|}$ be a signal,
  and $\varphi$ be an $\STL$ formula.
  We define the \emph{robustness} 
  $\Robust{\bw}{\varphi} \in \R \cup \{\infty,-\infty\}$ as follows, by induction.
  Here $\bigsqcap$ and $\bigsqcup$ denote infimums and supremums of real numbers, respectively. Their binary version $\sqcap$ and $\sqcup$ denote minimum and maximum.
\begin{displaymath}
\begin{array}{l}
\Robust{\bw}{f(x_1, \cdots, x_n) > 0}  \;\Defeq \;
f\bigl(\bw(0)(x_1), \cdots, \bw(0)(x_n)\bigr) 
\\
\Robust{\bw}{\bot}  \;\Defeq\;  -\infty
\qquad
\Robust{\bw}{\neg \varphi}   \;\Defeq\;   - \Robust{\bw}{\varphi}\qquad \\
          \Robust{\bw}{\varphi_1 \wedge \varphi_2}   \;\Defeq\;   \Robust{\bw}{\varphi_1} \sqcap \Robust{\bw}{\varphi_2}\\
          \Robust{\bw}{\varphi_1 \UntilOp{I} \varphi_2}   \;\Defeq\; 
                                                         \textstyle{ \Vee{t \in I}\bigl(\,\Robust{\bw^t}{\varphi_2} \sqcap 
                                                         \Wedge{t' \in [0, t)} \Robust{\bw^{t'}}{\varphi_1}\,\bigr)}
      \end{array}
    \end{displaymath}
\end{mydefinition}

Here are some intuitions and consequences of the definition. 
The robustness $\Robust{\bw}{f(\vec{x})>c}$ stands for the vertical margin $f(\vec{x})-c$ for the signal $\bw$ at time $0$.  A negative robustness value indicates how far the formula is from being true.
The robustness for the eventually modality is computed by
\begin{math}
\Robust{\bw}{\DiaOp{[a,b]} (x > 0)}
      = \Vee{t \in [a,b]} \bw(t)(x)
\end{math}.

The original semantics of $\STL$ is Boolean, given by a binary relation $\models$ between signals and formulas. The robust semantics refines the Boolean one as follows:
$ \sem{\bw,\varphi} > 0$
 implies
$\bw\models\varphi$, and
$ \sem{\bw,\varphi} < 0$
implies
$\bw\not\models\varphi$, see~\cite[Prop.~16]{FainekosP09}.
Optimization-based falsification via robust semantics hinges on this refinement. 
Although the definitions so far are for time-unbounded signals only,
we note that the robust semantics $\sem{\bw,\varphi}$,
as well as the Boolean satisfaction $\bw\models\varphi$, can be easily adapted to time-bounded signals (Def.~\ref{def:operationsOnSignals}).

Finally, here is a formalization of the falsification problem. It refines the description in~\S{}\ref{sec:introduction}. In particular, its use of real-valued robust semantics enables hill-climbing optimization. See Fig.~\ref{fig:fromQualitativeToQuantitative}. 
\begin{mydefinition}[falsifying input]\label{def:falsifyingInput}
Let $\mathcal{M}$ be a system model, and~$\varphi$ be an STL formula. A signal $\bu\colon [0,T]\to \Rin$ is a \emph{falsifying input} if $\Robust{\mathcal{M}(\bu)}{\varphi}<0$ (implying $\mathcal{M}(\bu)\not\models\varphi$).
\end{mydefinition}

\section{Two-Layered Optimization Framework with Monte Carlo Tree Search}
\label{sec:twoLayered}

In this section we present our main contribution, namely a two-layered
 optimization framework for hybrid system falsification. It combines:
 \emph{Monte Carlo tree search}
 (MCTS)~\cite{BrownePWLCRTPSC12} for
 high-level planning in the upper layer; and hill-climbing optimization (such as SA, GNM~\cite{LuersonLeRiche2004} and
 CMA-ES~\cite{AugerH05}) for local input search in the lower layer.
 See Fig.~\ref{fig:twoLayerd} for a schematic overview.
 The upper layer 
 steers the lower layer using the UCT strategy~\cite{KocsisS06}, an established method in  machine learning for balancing exploration and exploitation.
 
\begin{algorithm*}
\caption{\emph{Basic} Two-Layered Algorithm}
\label{algo:main1}
\begin{algorithmic}[1]
 \Require a system model $\mathcal{M}$,
 an $\STL$ formula $\varphi$, intervals $I_{i}=[u_{i}^{\min}, u_{i}^{\max}]$, $i\in\{1,\dotsc,M\}$, for the ranges of input $u_{1},\dotsc, u_{M}$ of $\mathcal{M}$,
 time horizon $T\in\Rpos$, and the following tunable parameters:  the number $K$ of control points,
 the number $L_{i}$ of partitions of the input range 
$[u_{i}^{\min} , u_{i}^{\max}]$ for each $i\in\{1,\dotsc,M\}$, the
 scalar $c$ in Line~\ref{line:ucb} of Alg.~\ref{algo:aux}, and 
 an MCTS budget (the maximum number of MCTS samples, Line~\ref{line:main1while})

\Statex
\Function{MCTSPreprocess}{}
\State $A\gets \{1,\dotsc,L_{1}\}\times\cdots\times \{1,\dotsc,L_{M}\}$
   \label{line:main1theSetOfActions}   \Comment{the set of actions}
 \State $\mathcal{T}\gets\{\varepsilon\}$
   \label{line:main1treeInit}
  \Comment{the MCTS search tree, initially  root-only}
 \State $N\gets (\varepsilon\mapsto 0)$ 
   \Comment{ visit count $N$  initialized, defined only for $\varepsilon$}
 \State $R\gets (\varepsilon\mapsto \infty)$ 
   \Comment{ reward function $R$  initialized}
 \State $\overrightarrow{\bu}\gets $ null
   \Comment{place holder for a falsifying input}
    \State $R_{\min} \gets \infty$
   \Comment{place holder for a minimum reward}
   \State $\overrightarrow{a}_{\min} \gets$ null
    \Comment{ the most promising action sequence}

	\While{$R(\varepsilon)\ge 0$ and within the MCTS budget}
              \label{line:main1while}
              \State \Call{MCTSSample}{$\varepsilon$}
              \label{line:main1callMCTSSample}
	\EndWhile
        \If{$\overrightarrow{\bu}\neq$ null}
                        \Comment{ a falsifying input is found already in preprocessing} \label{line:main1falsifyingSignalAlreadyFound}
          \State \Return $\overrightarrow{\bu}$
        \Else \Comment{return the most promising action sequence}
               \label{line:main1falsifyingSignalNotFoundInPreprocessing}
          \State\Return $\overrightarrow{a}_{\min}$
        \EndIf
	
\EndFunction

\Statex {}
\Function{MCTSSample}{$w$}
 \Comment{let $w=a_{1}\dotsc a_{d}$ with $a_{i}\in A$}
 \State $N(w)\gets N(w)+1$ 
 \If{$|w|<K$} 
   \If{$wa'\in\mathcal{T}$ for all $a'\in A$}
      \label{line:main1NonProgressive}
    \Comment{if all children have been expanded}
     \State{$a\gets \Call{UCBSample}{w}$}
       \label{line:main1callUCBSample}
       \Comment{pick a child $wa$ by UCB}
    \State{$\Call{MCTSSample}{wa}$} 
       \label{line:main1reccallMCTSSample}
       \Comment{recursive call}
     \State{$R(w)\gets \min_{a'\in A}R(wa')$} 
       \label{line:main1backPropagation}
       \Comment{back-propagation}
   \Else
     \State randomly sample $a\in A$ from $\{a\mid wa\not\in\mathcal{T}\}$
       \label{line:main1chooseAChildToExpand}
     \Comment{expand a random unexpanded child $wa$}
     \State $\mathcal{T}\gets\mathcal{T}\cup\{wa\}$
       \label{line:main1addTheChild}
 \State \label{line:main1playOutHillClimb}
       $\bu_{1},\dotsc,\bu_{K}$
 $\gets\argminHillClimb_{\substack{
 \bu_{1}\in\Call{Reg}{a_{1}},
 \dotsc,
 \bu_{d}\in\Call{Reg}{a_{d}},
 \\
 \bu_{d+1}\in\Call{Reg}{a},
 \\
\bu_{d+2},\dotsc,\bu_{K}\in
 I_{1}\times\cdots\times
 I_{M}}}
\sem{\mathcal{M}(\bu_{1}\dotsc\bu_{K}),
 \,\varphi} 
$\Comment{playout by hill-climbing}
\State $N(wa)\gets 0$\qquad
 \State
 $R(wa)\gets\sem{\mathcal{M}(\bu_{1}\dotsc
\bu_{K}), \,\varphi}$
    \label{line:main1rewardOfNewChild}
 \If{$R(wa)<0$} 
   \State
    $\overrightarrow{\bu}\gets\bu_{1}\dotsc
    \bu_{K}$\label{line:main1falsified}
    \Comment{a falsifying input is found and stored in $\overrightarrow{\bu}$}
 \EndIf
\If{$R(wa)<R_{\min}$} 
   \State $R_{\min} \gets R(wa)$ \qquad
  \State
 $\overrightarrow{a}_{\min}\gets a_1\dotsc a_d a$
 \EndIf
     \State {$R(w)\gets \min_{a'\in A}R(wa')$} 
       \label{line:main1backPropagation2}
       \Comment{back-propagation}
   \EndIf
 \EndIf
\EndFunction

\Statex{}
\Function{Main}{}
\State $\overrightarrow{x}\gets$ \Call{MCTSPreprocess}{}
\label{line:main1Preprocess}
\If{$\overrightarrow{x}=\overrightarrow{\bu}$, an input signal} 
 \Comment{Line~\ref{line:main1falsifyingSignalAlreadyFound}}
   \State \Return $\overrightarrow{\bu}$\label{line:main1SimplyReturnTheResultOfPreprocessingPhase}
   \Else
  \Comment{$\overrightarrow{x}=a_{1}a_{2}\dotsc a_{K'}\in A^{*}$ with some $K'\le K$, Line~\ref{line:main1falsifyingSignalNotFoundInPreprocessing}}
\State  \Return $\argminHillClimb_{\substack{\bu_{1}\in \Call{Reg}{a_1}, \dots,\bu_{K'} \in \Call{Reg}{a_{K'}},\\\bu_{K'+1},\dotsc,\bu_{K}\in I_{1}\times\cdots\times I_{M} }}
  \sem{\mathcal{M}(\bu_{1}\dotsc \bu_{K}), \,\varphi}$
   \label{line:main1SecondHillClimb}
\EndIf
\EndFunction
\end{algorithmic}
\end{algorithm*}

\begin{algorithm*}[tbp]
\caption{Auxiliary Functions for Algs.~\ref{algo:main1} \&~\ref{algo:main2}}
\label{algo:aux}
\begin{algorithmic}[1]

\Statex {}
\Function{UCBSample}{$w$}
\State \Return $\argmax_{a\in A} \left(\bigl(1-\displaystyle\frac{R(wa)}{\max_{w'\in\mathcal{T}}R(w')}\bigr) + c\sqrt{\displaystyle\frac{2\ln N(w)}{N(wa)}}\;\right)$
   \label{line:ucb}
\EndFunction

\Function{Reg}{$a$} 
\Comment{The input region for an action $a\in A$ is of the form $(k_{1},\dotsc, k_{M})$, see Line~\ref{line:main1theSetOfActions} of Alg.~\ref{algo:main1}}
\State \Return
 $\prod_{i=1}^{M}\Bigl[\,
  u_{i}^{\min} + \frac{k_{i}-1}{L_{i}}(u_{i}^{\max}-u_{i}^{\min})\,,\;$
$
  u_{i}^{\min} + \frac{k_{i}}{L_{i}}(u_{i}^{\max}-u_{i}^{\min})
\,\Bigr]
$
\EndFunction

\end{algorithmic}
\end{algorithm*}

\begin{algorithm*}[tbp]
\caption{Two-Layered Algorithm with \emph{Progressive Widening}}
\label{algo:main2}
\begin{algorithmic}[1]
 \Require The same data as required in Alg.~\ref{algo:main1}, and additionally, constants
 $C,\alpha$ (used in Line~\ref{line:main2progressiveWidening})

\Statex

\Statex The algorithm is the same as Alg.~\ref{algo:main1}, except that the function $\Call{MCTSSample}{}$ is replaced by the following one.

\Statex {}
\Function{MCTSSample}{$w$}
 \Comment{let $w=a_{1}\dotsc a_{d}$ with $a_{i}\in A$}
\State $N(w)\gets N(w)+1$ 
 \If{$|w|<K$}
   \If{$
   \left(
     \begin{array}{l}
      \bigl|\{a'\in A\mid wa'\in \mathcal{T}\}\bigr|\ge C\cdot N(w)^{\alpha}
       \\
      \text{or $wa'\in\mathcal{T}$ for all $a'\in A$}
     \end{array}
 \right)
   $ } \label{line:main2progressiveWidening}
    \Comment{progressive widening: all or enough children  expanded}
     \State{$a\gets \Call{UCBSample}{w}$}
       \label{line:main2callUCBSample}
       \Comment{pick a child $wa$ by UCB}
    \State{$\Call{MCTSSample}{wa}$} 
       \label{line:main2reccallMCTSSample}
       \Comment{recursive call}
     \State{$R(w)\gets \min_{a'\in A}R(wa')$} 
       \label{line:main2backPropagation}
       \Comment{back-propagation}
   \Else
     \State $S\gets \text{(a maximal convex subset of $\bigcup_{wa'\not\in\mathcal{T}} \Call{Reg}{a'})$}$
     \label{line:main2convexSubset}
     \State \label{line:main2HillClimbForExpansion}
$\bu_{1},\dotsc,
\bu_{K}$    
$\gets\argminHillClimb_{\substack{
 \bu_{1}\in\Call{Reg}{a_{1}},
 \dotsc,
 \bu_{d}\in\Call{Reg}{a_{d}},
 \\
 \bu_{d+1}\in S,
 \\
\bu_{d+2},\dotsc,\bu_{K}\in
 I_{1}\times\cdots\times
 I_{M}}}$
$\sem{\mathcal{M}(\bu_{1}\dotsc\bu_{K}),
 \,\varphi}$\Comment{playout by hill-climbing}

\State 
 $a\gets\bigl(\text{$a\in A$ such that $\bu_{d+1}\in\Call{Reg}{a}$}\bigr)$
\label{line:main2ChooseChild}
     \State $\mathcal{T}\gets\mathcal{T}\cup\{wa\}$
       \label{line:main2addTheChild}
 \State $N(wa)\gets 0$\qquad
 \State
 $R(wa)\gets\sem{\mathcal{M}(\bu_{1}\dotsc
\bu_{K}), \,\varphi}$
    \label{line:main2rewardOfNewChild}
  \If{$R(wa)<0$} 
   \State
    $\overrightarrow{\bu}\gets\bu_{1}\dotsc
    \bu_{K}$ \label{line:main2falsified}
\EndIf
   \If{$R(wa)<R_{\min}$} 
   \State $R_{\min} \gets R(wa)$\qquad
   \State
 $\overrightarrow{a}_{\min}\gets a_1\dotsc a_d a$
 \EndIf
      \State{$R(w)\gets \min_{a'\in A}R(wa')$} 
       \label{line:main2backPropagation2}
       \Comment{back-propagation}

   \EndIf
 \EndIf
\EndFunction

\end{algorithmic}
\end{algorithm*}

\begin{figure}[tbp]
 \centering
   \includegraphics[width=\columnwidth]{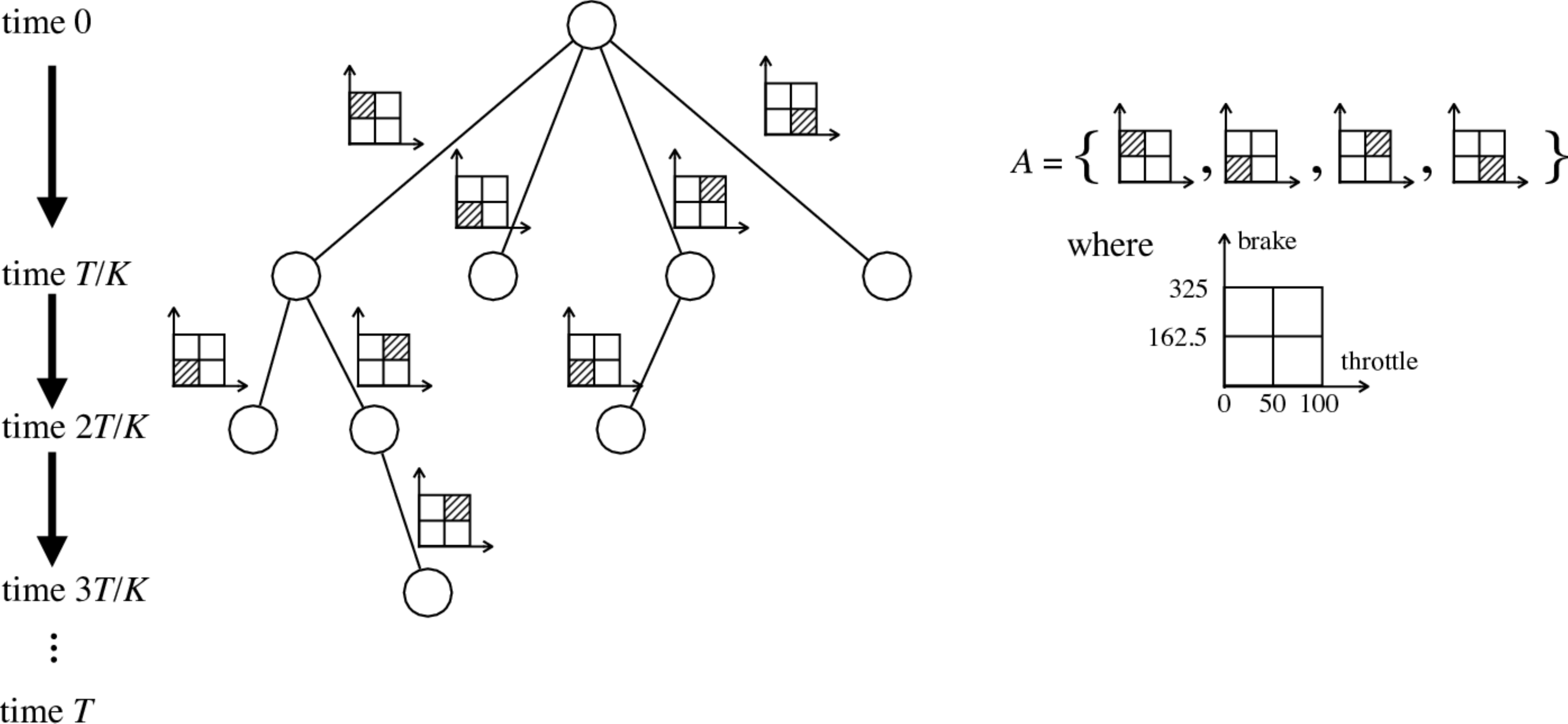}
 \caption{Our MCTS search tree for a system model $\mathcal{M}$ with two input signals, throttle and brake, whose ranges are $[0,100]$ and $[0,325]$, respectively. We partition each range into two intervals, i.e.\ $L_{1}=L_{2}=2$, hence the branching degree~$|A|$ is $2\times 2$. }
 \label{fig:mctsTree}
\end{figure}

We present two algorithms: the \emph{basic} two-layered algorithm (Alg.~\ref{algo:main1}), and a version enhanced with \emph{progressive widening} (Alg.~\ref{algo:main2}). 
The auxiliary functions used therein are presented in Alg.~\ref{algo:aux}. Our algorithms work on an \emph{MCTS search tree}, as illustrated in Fig.~\ref{fig:mctsTree}. 

\subsection{The Basic Two-Layered Algorithm (Alg.~\ref{algo:main1})}
\label{subsec:algoMain1}
We start with Alg.~\ref{algo:main1}, using the example in Fig.~\ref{fig:mctsTree}.

\paragraph{Time Staging} We search for a falsifying input signal, focusing on piecewise-constant signals (Fig.~\ref{fig:timeStaging}, left). The interval $[0,T]$ is divided into $K$ equal sub-intervals ($K$ is a tunable parameter). The time points $0,\frac{T}{K},\frac{2T}{K},\dotsc,\frac{(K-1)T}{K}$ at which those intervals start are called \emph{control points}.   Our goal is therefore to find a sequence $\bu_{1},\dotsc,\bu_{K}$, where each $\bu_{i}=(u_{i1},\dotsc,u_{iM})$ is an $M$-dimensional real vector ($M$ is the number of input signal dimensions for the model $\mathcal{M}$), so that the corresponding piecewise-constant signal is a falsifying one (Def.~\ref{def:falsifyingInput}).

 We assume
 intervals $I_{i}=[u_{i}^{\min}, u_{i}^{\max}]$, $i\in\{1,\dotsc,M\}$, for the ranges of input $u_{1},\dotsc, u_{M}$ of the model $\mathcal{M}$.
 
 \begin{figure*}[tbp]
 \centering
   \includegraphics[width=\textwidth]{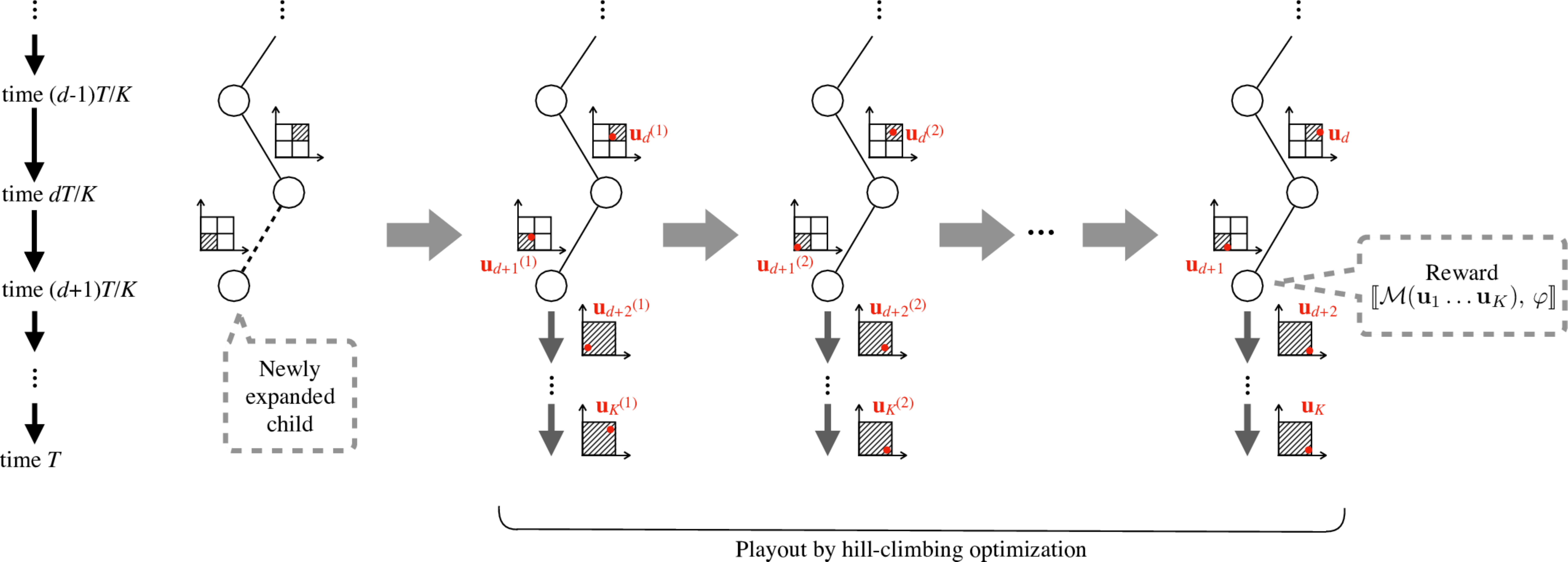}
 \caption{Playout by hill-climbing optimization}
 \label{fig:playout}
\end{figure*}

\paragraph{The Search Tree} A search tree in MCTS has a branching degree~$|A|$, where the set $A$ is  called an \emph{action set} in the MCTS literature. In Go, for example, an action set $A$ consists of possible moves.

We use, as the action set $A$, a \emph{partitioning} of the input space $I_{1}\times\cdots\times I_{M}$. We partition the input space into $L_{1}\times\cdots \times L_{M}$ hypercubes of equal size, according to predetermined parameters $L_{1},\dotsc, L_{M}$, where $M$ is the number of input signals of the system and $L_i$ indicates how finely the $i$-th input should be partitioned.
 In Fig.~\ref{fig:mctsTree} we present an example where $M=2$ and $L_{1}=L_{2}=2$. There we have four actions in the set $A$, corresponding to the four square regions.

An edge in our search tree represents a choice of an input region---from which we choose the input value $\bu_{i}$---for a single control point $\frac{(i-1)T}{K}$. The depth of the tree is $K$ (the number of control points).
We follow the usual  convention and specify a node of a $|A|$-branching tree by a word $w=a_{1}a_{2}\dotsc a_{j}$ over the alphabet $A$, where $j\le K$. That is: the root is $\varepsilon$ (the empty word), its child in the direction $a_{1}\in A$ is $a_{1}$, its children are $a_{1}a_{1}, a_{1}a_{2},\dotsc$, and so on.

In general, a node in an MCTS search tree is decorated by two values: \emph{reward} $R$ and \emph{visit count} $N$.
In our case, $R$ stores the current estimate of the smallest (i.e.\ the best) robustness value.
Both values are updated explicitly during back-propagation (see below).

\paragraph{Monte Carlo Tree Search Sampling}
Much like usual MCTS, Alg.~\ref{algo:main1} iteratively expands the
search tree $\mathcal{T}$. Initially the tree $\mathcal{T}$ is root-only
(Line~\ref{line:main1treeInit}), and in each iteration---called
\emph{MCTS sampling}---the invocation of $\Call{MCTSSample}{}$ on Line~\ref{line:main1callMCTSSample} adds one
new node to~$\mathcal{T}$. In the MCTS literature, \emph{expanding} a
child means adding the child to $\mathcal{T}$. We repeat MCTS sampling
until a counterexample is found, or the MCTS budget is used up
after the maximum number of iterations (Line~\ref{line:main1while}).

The exploration-exploitation trade-off in MCTS comes in the choice of the node to add. In each MCTS sampling, we start  from the root (Line~\ref{line:main1callMCTSSample}), walk down the tree $\mathcal{T}$, choosing already expanded nodes (Lines~\ref{line:main1callUCBSample}--\ref{line:main1reccallMCTSSample}), until we expand a  child (Lines~\ref{line:main1chooseAChildToExpand}--\ref{line:main1addTheChild}).
Growing a wider tree means exploration, while a deeper tree means exploitation.

We use the UCT strategy~\cite{KocsisS06}, the most commonly used strategy in MCTS, to resolve the dilemma. UCT is based on the UCB strategy for the multi-armed bandit problems~\cite{Auer2002};  Line~\ref{line:ucb} of Alg.~\ref{algo:aux} follows UCB, where the exploitation score $1-\frac{R(wa)}{\max_{w'\in\mathcal{T}}R(w')}$ and the exploration score $\sqrt{\frac{2\ln N(w)}{N(wa)}}$ are superposed using a scalar $c$. Recall that our rewards $R(wa)$ for $w$'s children are given by robustness estimates from previous simulations, and that falsification favors smaller $R$.  Note also that values of $R$ can be greater than $1$. In the exploitation score $1-\frac{R(wa)}{\max_{w'\in\mathcal{T}}R(w')}$, therefore, we normalize rewards to the interval $[0,1]$ and reverse their order.\footnote{We can assume nonnegative values of $R$ , otherwise we already have a falsifying input.} The exploration score  $\sqrt{\frac{2\ln N(w)}{N(wa)}}$ is taken from UCB: the visit count $N(w)$ gives how many times the node $w$ has been visited, that is, how many  offspring the node $w$ currently has in~$\mathcal{T}$. The scalar, for the trade-off, is a tunable parameter, as usual in MCTS. 

\paragraph{Playout and Back-Propagation} In MCTS, the reward of a newly expanded node $a_{1}a_{2}\dotsc a_{d}a$ (see e.g.\  Line~\ref{line:main1addTheChild}) is computed by an operation called \emph{playout}. The result is then \emph{back-propagated}, in a suitable manner, to the ancestors: $a_{1}\dotsc a_{d}$, $a_{1}\dotsc a_{d-1}$, $\dotsc$, and finally $\varepsilon$.

In our MCTS algorithms for falsification we use hill-climbing optimization (e.g. SA, GNM and 
 CMA-ES) for playout. See Line~\ref{line:main1playOutHillClimb}, where input values $\bu_{1},\bu_{2},\dotsc,\bu_{K}$ are sampled by stochastic hill-climbing optimization, so that the resulting robustness value of the specification $\varphi$ becomes smaller. The regions from which to sample those values are dictated by the MCTS tree:
  $\bu_{1}\in\Call{Reg}{a_{1}},
 \dotsc,
 \bu_{d}\in\Call{Reg}{a_{d}}$ follow the actions $a_{1},\dotsc,a_{d}$ determined so far (here $\Call{Reg}{}$ is from Alg.~\ref{algo:aux}); $\bu_{d+1}\in\Call{Reg}{a}$ follows the newly chosen action $a$ (Line~\ref{line:main1chooseAChildToExpand}); and the remaining values $\bu_{d+2},\dotsc,\bu_{K}$ can be chosen from the whole input range $ I_{1}\times\cdots\times
 I_{M}$.

\figurename~\ref{fig:playout} illustrates an example of playout by hill-climbing optimization.  Smaller gray squares represent actions, and red dots represent input values (notice that they are chosen from the gray regions). The values $\bu_{1},\dotsc,\bu_{K}$ are sampled repeatedly so that the robustness value $\sem{\mathcal{M}(\bu_{1}\dotsc\bu_{K}), \,\varphi}$ becomes smaller.

An intuition of this playout operation is that we sample the best input signal, $\bu_{1}\dotsc \bu_{K}$, under the constraints  imposed by the MCTS search tree (namely, the input regions prescribed by the actions). The least robustness value thus obtained is assigned to the newly expanded  node $wa$ as its reward (Line~\ref{line:main1rewardOfNewChild}). If $R(wa)<0$ then this means we have already succeeded in falsification (Line~\ref{line:main1falsified}).

\emph{Back-propagation} is an important operation in MCTS. Following the intuition that the reward $R(w)$ is the smallest robustness achievable at the node $w$, we define the reward of an internal node~$w$ by the minimum of its children's rewards. See Lines~\ref{line:main1backPropagation} and~\ref{line:main1backPropagation2}. Note that, via recursive calls of $\Call{MCTSSample}{}$ (Line~\ref{line:main1reccallMCTSSample}), the result of playout is propagated to all ancestors.

\paragraph{A Two-Layered Framework} 
In Alg.~\ref{algo:main1}, hill-climbing optimization occurs twice, in Lines~\ref{line:main1playOutHillClimb} and~\ref{line:main1SecondHillClimb}. The first occurrence is in playout of MCTS---this way we interleave  MCTS optimization  (by growing a tree) and hill-climbing optimization. See Fig.~\ref{fig:twoLayerd}.  MCTS optimization is considered to be a \emph{preprocessing} phase in Alg.~\ref{algo:main1} (Line~\ref{line:main1Preprocess}): its principal role is to find an action sequence $\overrightarrow{a}_{\min}$, i.e.\ a sequence of input regions, that is most promising. In the remainder of the $\Call{Main}{}$ function, the second hill-climbing optimization is conducted for falsification, where we sample according to  $\overrightarrow{a}_{\min}$. 

The two occurrences of hill-climbing optimization therefore have different
roles. Given also the fact that the first occurrence is repeated every
time we expand a new child, we choose to spend less time for the former
than the latter. 
In our implementation, we set the timeout to be 5--15 seconds
for the first 
hill-climbing sampling in Line~\ref{line:main1playOutHillClimb}
($\text{TO}_\text{po}$
in~\S{}\ref{sec:experiments}), while for the second hill-climbing
sampling in Line~\ref{line:main1SecondHillClimb} the timeout is 300 seconds.

A falsifying input $\overrightarrow{\bu}$ is often found already in the preprocessing phase. In this case the $\Call{Main}{}$ function simply returns $\overrightarrow{\bu}$ (Line~\ref{line:main1SimplyReturnTheResultOfPreprocessingPhase}).

\subsection{The Two-Layered Algorithm with Progressive Widening (Alg.~\ref{algo:main2})}
Our second algorithm (Alg.~\ref{algo:main2}) differs from the basic one (Alg.~\ref{algo:main1}) in two ways:

\paragraph{Progressive Widening}
Alg.~\ref{algo:main2} uses \emph{progressive widening}~\cite{Coulom07}; see Line~\ref{line:main2progressiveWidening}. Unlike in the basic algorithm (Line~\ref{line:main1NonProgressive} of Alg.~\ref{algo:main1}), we do not always expand a new child, even if there are unexpanded ones; the threshold $C\cdot N(w)^{\alpha}$ is computed using the visit count $N(w)$ and tunable parameters $C,\alpha$.

Progressive widening is a widely employed technique in MCTS for coping with a large or infinite action set $A$---in such a case expanding all children incurs a lot of computational cost. See e.g.~\cite{lee2015adaptive}. In our Alg.~\ref{algo:main2} the action set $A$ can be large, depending on the numbers $L_{1},\dotsc,L_{m}$ of input range partitions.

\paragraph{Hill-Climbing Optimization for Expanding Children}
In progressive widening, since we may not expand all the children, it makes sense to be selective about which child to expand. This is in contrast to random sampling  in Alg.~\ref{algo:main1} (Line~\ref{line:main1chooseAChildToExpand}). See Line~\ref{line:main2HillClimbForExpansion} of Alg.~\ref{algo:main2}, where we first playout by hill-climbing optimization. The value $\bu_{d+1}$ thus obtained is then used to determine which child $wa$ to expand, in Line~\ref{line:main2ChooseChild}. In order to ensure that the new child $wa$ is indeed previously unexpanded, the value $\bu_{d+1}$ is sampled from the set $\bigcup_{wa'\not\in\mathcal{T}} \Call{Reg}{a'}$; in fact, we
\begin{figure}
 \centering
      \includegraphics[width=0.32\textwidth]{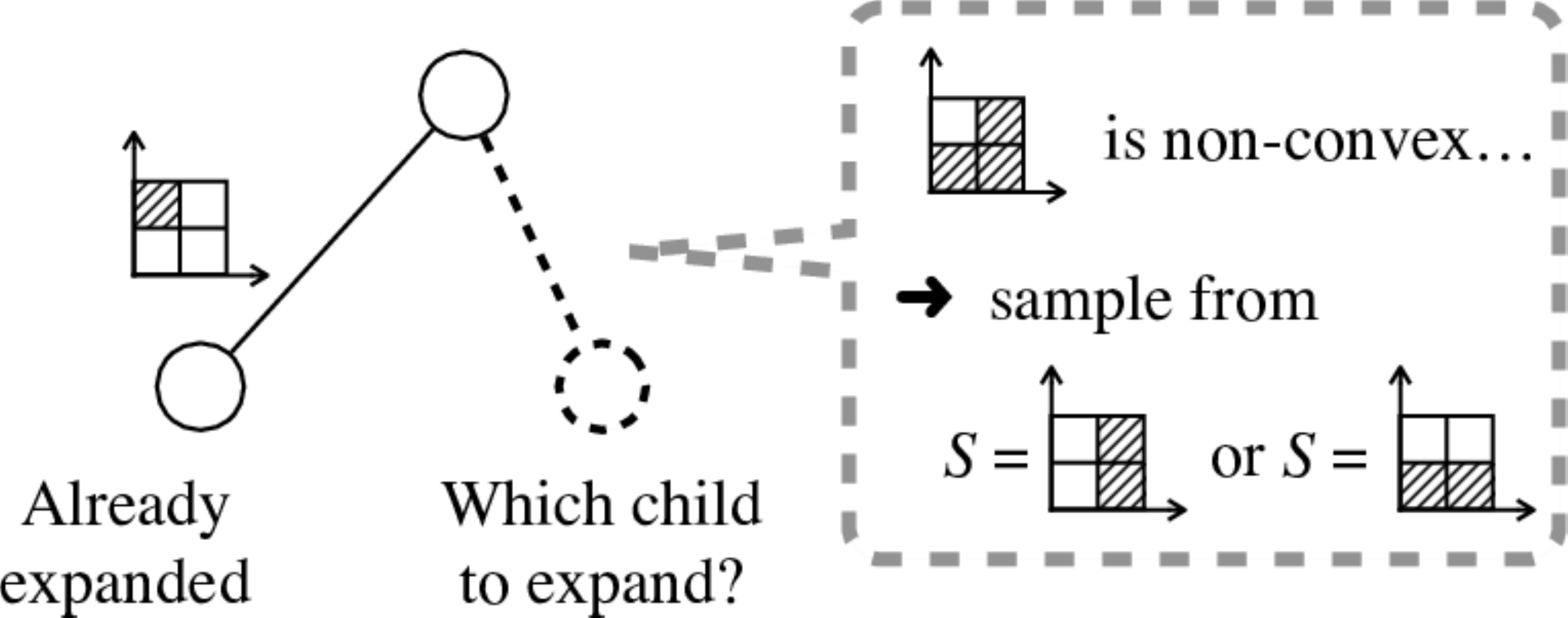}
    \caption{Lines~\ref{line:main2convexSubset}--\ref{line:main2ChooseChild} of Alg.~\ref{algo:main2}}
   \label{fig:convexSubset}
\end{figure}
restrict to its convex subset (Line~\ref{line:main2convexSubset}), because many hill-climbing optimization algorithms work best in a convex domain. See Fig.~\ref{fig:convexSubset} for illustration.

\subsection{Discussion}
Our algorithms interleave  MCTS optimization and hill-climbing optimization: the latter is used in the playout operation of the former, for sampling and estimating the reward of a high-level input-synthesis strategy. This high-level strategy is concretely given by a sequence $a_{1}a_{2}\dotsc a_{d}$ of input regions. Via the UCT tree search strategy, we ensure that our search in a search tree is driven not only by depth, but also by width.
This way we enhance exploration in search-based falsification, in the sense that different regions of the input space are sampled in a structured and disciplined manner. 
It is an interesting topic for future work to quantify the \emph{coverage guarantees} that can potentially be achieved by our approach.

In falsification of hybrid systems, it is often the case   that \emph{simulation}, i.e.\ running a model $\mathcal{M}$ under a given input signal, is  computationally the most expensive operation. In our algorithm this occurs in
Lines~\ref{line:main1playOutHillClimb} and~\ref{line:main1SecondHillClimb}, since a hill-climbing optimization algorithm  tries many samples of $\bu_{1},
\dotsc,\bu_{K}$. Simplifying Line~\ref{line:main1playOutHillClimb}, e.g.\ by decimating the control points, can result in a useful variation of our algorithm.

Among the tunable parameters of the algorithm is the scalar $c$, used for the UCB sampling (Line~\ref{line:ucb} of Alg.~\ref{algo:aux}). Having this parameter is unique to our falsification framework, in comparison to simple robustness-guided optimization (with hill-climbing only). Specifically,  the parameter $c$ endows our algorithm with \emph{flexibility} in the exploration-exploitation trade-off. Given the diversity of instances of the hybrid system falsification problem, it is unlikely that there is a single value of $c$ that is optimal for all falsification examples. An engineer can then use her/his expert domain knowledge to tune the parameter $c$.

\section{Experimental Evaluation}\label{sec:experiments}
We have implemented our \emph{basic} algorithm (Alg.~\ref{algo:main1}, denoted ``Basic'') and our \emph{progressive widening} algorithm (Alg.~\ref{algo:main2}, denoted ``P.W.'') in MATLAB,
using Breach~\cite{Donze10} as a front-end for hill-climbing optimization
and for its implementation of the robust semantics.

The experiments have two goals. Firstly, in~\S{}\ref{sec:performance}, we evaluate the falsification performance of our proposal
in comparison to the state-of-the-art. Since our MCTS enhancement emphasizes coverage, our interest is in the success rate in hard problem instances rather than in execution time. Secondly, in~\S{}\ref{sec:parameters}, we evaluate the impact of different choices of parameters for our
algorithms (such as the UCB scalar~$c$ in  Alg.~\ref{algo:aux}).

\subsection{Experiment Setup}

The experiments are based on the following benchmarks.

The \emph{automatic transmission}  (AT) model is a Simulink model that
was proposed as a benchmark for falsification
in~\cite{HoxhaAF14}.
It has input signals~$\throttle \in [0,100]$ and~$\brake \in [0,325]$,
and computes the car's speed $\speed$, the engine rotation~$\rpm$,
and the selected gear~$\gear$.
We consider the following specifications, taken in part from~\cite{HoxhaAF14}.

S1${} \equiv \Box_{[0,30]}~(\speed < 120)$ can be falsified
easily by hill-climbing with an input $\throttle = 100$ and $\brake = 0$ throughout.

S2${} \equiv \Box_{[0,30]}~(\gear = 3 \to \speed \ge 20)$ states that in gear three,
the speed should not get too low. The difficulty arises from the lack of guidance by robustness as long as $\gear \neq 3$: we follow~\cite{HoxhaAF14} and take $\gear=1,\dotsc,\gear=4$ as Boolean propositions, instead of taking $\gear$ as a numeric variable. In contrast to \cite{HoxhaAF14}, we use a more difficult speed threshold of 20 instead of 30.

S3${} \equiv \Diamond_{[10,30]}~(\speed \not\in [53,57])$ states that it is not possible to maintain a constant speed after 10s. A falsifying trace needs precise inputs to hit and maintain the narrow speed range.

S4${} \equiv \Box_{[0,29]}(\speed<100) \lor \Box_{[29,30]}(\speed>65)$
is a specification designed to demonstrate 
the limitation of  robustness-guided falsification by hill-climbing optimization only. Here, a falsifying trajectory has to reach high speed
before braking. Similarly to~S2, the speed~100 has to be reached
much earlier than the indicated time bound of~29 to give sufficient time for
deceleration. However, by using the maximum as semantics for the $\lor$-connective, the robustness computation can shadow either of
the disjuncts .

S5${} \equiv \Box_{[0,30]}(\rpm < 4770 \lor \Box_{[0,1]}(\rpm > 600))$ aims to
prevent systematic sudden drops from high to low $\rpm$. It is falsified if an $\rpm$ peak above 4770 is immediately followed by a drop to $\rpm \le 600$.

The second benchmark is the \emph{Abstract Fuel Control} (AFC) mod\-el~\cite{JinDKUB14}.
It takes two input signals,
\emph{pedal angle} and \emph{engine speed}, and outputs the critical signal
\emph{air-fuel ratio} ($\AF$), which influences fuel efficiency and car performance.
The value is expected to be close to a reference value $\AFref$.
The pedal angle varies in the range~$[0,61.1]$
and the engine speed varies in the range~$[900,1100]$.
According to~\cite{JinDKUB14},
this setting corresponds to \emph{normal mode}, where $\AFref = 14.7$.

The basic requirement of the AFC is to keep the air-to-fuel ratio $\AF$ close
to the reference $\AFref$. However, changes to the pedal angle cause brief spikes in the
output signal~$\AF$ before the controller is able to regulate the engine.
Falsification is used to discover the amplitude and periods of such spikes.

The formal specification Sbasic is $\BoxOp{[11,30]}(\neg(|\AF-\AFref|>0.05*14.7))$. It is violated when $\AF$ deviates from its $\AFref$ too much.
Another specification is Sstable:
$\neg(\DiaOp{[6,26]}\BoxOp{[0,4]}(|\AF - \AFref| > 0.01 * 14.7))$.
The goal is to find spikes where the ratio is off by a fraction $0.01$ of the reference value for at least $t'$ seconds during the interval~$[6,26]$.

The third benchmark model is called \emph{Free Floating Robot} (FFR)
that has been considered as a falsification benchmark in~\cite{DeshmukhHJMP17}.
It is a robot vehicle powered by four boosters and moving in two spatial dimensions.
It is governed by the following second-order differential equations:
\begin{align*}
\ddot x    & =  0.1 \cdot (u_1 + u_3) \cos(\phi) - 0.1 \cdot (u_2 + u_4) \sin(\phi) \\
\ddot y    & =  0.1 \cdot (u_1 + u_3) \sin(\phi) + 0.1 \cdot (u_2 + u_4) \cos(\phi) \\
\ddot \phi & = 5/12 \cdot (u_1 + u_3) - 5/12 \cdot (u_2 + u_4)
\end{align*}
The goal of the robot is to steer from $(x,y,\phi) = (0,0,0)$ to $x = y = 4$, with a
tolerance of $0.1$, such that $\dot x$ and $\dot y$ are within $[-1,1]$,
given a time horizon of $T = 5$.
The four inputs $u_i \in [-10,10]$ range over the same domain.
We run falsification on the negated requirement:
Strap${} \equiv \lnot\ \Diamond_{[0,5]}\ x,y \in [3.9,4.1] \land \dot x, \dot y \in [-1,1]$.

The experiments use Breach version 1.2.9 and MATLAB~R2017b on an Amazon EC2
c4.large instance (March 2018, 2.9~GHz Intel Xeon E5-2666, 2 virtual CPU cores, 4~GB main memory).

\subsection{Performance Evaluation}
\label{sec:performance}

\begin{table*}
\setlength\tabcolsep{2.5pt}
\caption{Comparison of uniform random sampling and Breach against
Algs.~\ref{algo:main1} (Basic) and \ref{algo:main2} (P.W.).
         \normalfont
For each specification, we show the success rate out of 10 trials,
and the average run time (in seconds) of those trials which were successful.
The respective parameters are shown in the leftmost columns:
M\_b (MCTS budget) is the maximum visit count for the root of the MCTS search tree (i.e.\ the maximum number of nodes of the tree); $\text{TO}_\text{po}$ (in seconds) is the  timeout (wall-clock)
for each individual MCTS playout by hill-climbing optimization;
$c$~is the scalar for the  exploration-exploitation trade-off in UCB (Alg.~\ref{algo:aux}). 
The number $K$ of control points is  $5$ for AT and AFC, and $3$ for FFR.
The partitioning $L$ of the input space w.r.t. each dimension is $3\times 5$ for AT (throttle, brake) and AFC (pedal, engine), and $2\times 2\times 2\times 2$ for FFR ($u_1, u_2, u_3, u_4$).
For progressive widening
(Alg.~\ref{algo:main2}) we use the parameters
$C = 0.7$ and $\alpha = 0.85$. 
Timeout for the hill-climbing in the end (Line~\ref{line:main1SecondHillClimb} of Alg.~\ref{algo:main1}) is 300 seconds.
For random testing, timeout is $900s$.
The cells with bold fonts are local best performers w.r.t. each hill-climbing solver, and green backgrounded cells are the global performers w.r.t. each property.
Here, the ranking criterion takes success rate as first priority, and average time as second priority.
}
\label{tab:results}
\small
\begin{tabular}{llrrrrrrrrrrrrrrrrrrrr}
\toprule
    &
    & \multicolumn{3}{l}{~Parameters}
    & \multicolumn{10}{l}{~AT model}
    & \multicolumn{4}{l}{~AFC model}
    &\multicolumn{2}{l}{~FFR model}
    \\
\cmidrule(lr){3-5}
\cmidrule(lr){6-15}
\cmidrule(l){16-19}
\cmidrule(l){20-21}
    &
    &
    &
    &
    & \multicolumn{2}{c}{S1}
    & \multicolumn{2}{c}{S2}
    & \multicolumn{2}{c}{S3}
    & \multicolumn{2}{c}{S4}
    & \multicolumn{2}{c}{S5}
    & \multicolumn{2}{c}{Sbasic}
    & \multicolumn{2}{c}{Sstable}
    & \multicolumn{2}{c}{Strap}
    \\
   \multicolumn{2}{c}{Algorithm}
    & M\_b & $\text{TO}_\text{po}$
    & \multicolumn{1}{c}{$c$}
    & succ. & time
    & succ. & time
    & succ. & time
    & succ. & time
    & succ. & time
    & succ. & time
    & succ. & time
    & succ. & time
    \\
\cmidrule(r){1-2}
\cmidrule(lr){3-5}
\cmidrule(lr){6-7}
\cmidrule(lr){8-9}
\cmidrule(lr){10-11}
\cmidrule(lr){12-13}
\cmidrule(lr){14-15}
\cmidrule(lr){16-17}
\cmidrule(lr){18-19}
\cmidrule(l){20-21}

 \multicolumn{2}{c}{Random}  &  &                 &      & 10/10 &  108.9 & 10/10 &   289.1 &      1/10 & 301.1 &  0/10 &  - &  0/10 &  -  &  6/10  & 278.7   &    \tbcolor  10/10 & \tbcolor 242.6  &4/10 & 409.3  \\\midrule
\multirow{3}{*}{\rotatebox{90}{\begin{footnotesize}CMA-ES\end{footnotesize}}}
& Breach &    &                 &      & 10/10 &  21.9 & 6/10 &   30.3 & \textbf{10/10} & \textbf{193.9} &   4/10 & 208.8 &  3/10 &  75.5  & \tbcolor \textbf{10/10}   &  \tbcolor\textbf{111.7}   &      3/10 &  256.3&\tbcolor\textbf{10/10} &\tbcolor\textbf{119.8} \\
& Basic   & 40 & 15 & 0.20 & 10/10 &  15.8 & 10/10 & 108.5 & 10/10 & 697.1  &  7/10 & 786.8 &  9/10 & 384.4   & 10/10& 182.0   &7/10&336.9& 10/10 &338.0 \\
& P.W.   & 40 & 15  & 0.20 & \textbf{10/10} &  \textbf{10.8} & \tbcolor\textbf{10/10} &\tbcolor \textbf{65.7} &  10/10 & 728.6 &  \textbf{7/10} & \textbf{767.8} &  \textbf{10/10} & \textbf{648.1}&  10/10 & 177.1 &\textbf{8/10}&\textbf{272.9}&10/10 &473.9\\
\midrule
\multirow{3}{*}{\rotatebox{90}{GNM}}
& Breach &    &                  &      & \tbcolor\textbf{10/10} &  \tbcolor\textbf{5.4} & \textbf{10/10} & \textbf{151.4} &  0/10 &     - &  0/10 &     - &  0/10 &     -
&  \textbf{10/10}& \textbf{171.4}      & 0/10&- & 0/10 & - \\
& Basic   & 20 &  5  & 0.20 & 10/10 &  12.4 & 10/10 & 162.3 &\tbcolor\textbf{10/10} &\tbcolor\textbf{185.6}  &  7/10 & 261.9 &  7/10 & 163.7&    10/10&  227.1     & 2/10&378.5& \textbf{10/10}&\textbf{162.2}\\
& P.W.   & 20 &  5  & 0.05 & 10/10 & 60.8 &  9/10 & 110.7 &  8/10 & 211.2  & \tbcolor \textbf{8/10} & \tbcolor\textbf{313.0} & \tbcolor\textbf{10/10} &\tbcolor\textbf{178.7} &  10/10& 252.0   &\textbf{6/10}&\textbf{153.2}& 6/10 &197.4\\
\midrule
\multirow{3}{*}{\rotatebox{90}{SA}}
& Breach &    &    &               & \textbf{10/10} & \textbf{160.1} &  0/10 &     - &  3/10 & 383.7 &  0/10 &     - &  3/10 &  80.4 &   0/10&  -    &6/10&307.0& 3/10 & 92.8\\
& Basic   & 20 & 15  & 0.05 & 10/10 & 264.8 &  9/10 & 236.1 & \textbf{8/10} & \textbf{385.6}  &  \textbf{8/10} & \textbf{505.3} &  7/10 & 341.2 &    5/10&391.3  &\textbf{8/10}& \textbf{273.8}& \textbf{10/10} &\textbf{273.2} \\
& P.W.   & 40 & 15 & 0.20 & 10/10 & 208.7 & \textbf{10/10} & \textbf{377.6} &  8/10 & 666.0  &  7/10 & 795.4 & \textbf{10/10} & \textbf{624.2}&   \textbf{8/10}   &  \textbf{665.7}     &6/10&293.7&10/10 &390.9\\
\bottomrule
\end{tabular}
\end{table*}

The results are shown in Table~\ref{tab:results} and are grouped with respect to the method:
uniform random sampling (``Random'') as a baseline, Breach, our ``Basic'' algorithm (Alg.~\ref{algo:main1}) and our ``P.W.'' algorithm (Alg.~\ref{algo:main2}), as well as with respect to the underlying hill-climbing optimization solver (CMA-ES, GNM and SA).
Run times are shown in seconds. Since the algorithms are
stochastic, we give the success rate out of a number of trials.

For all the experiments, input signals are chosen to be piecewise constant,
with $K = 5$ control points for AT and AFC, and $K = 3$ control
points for FFR (due to the shorter time horizon).
These numbers coincide with the depth of the MCTS search trees.
In Breach, this is achieved with the ``UniStep'' input generator,
with its \texttt{.cp} attribute set to $K$. The timeout for Breach was set to 900 seconds
(which is well above all successful falsification trials)
with no upper limit on the number of simulations.
For our P.W.\ algorithm, we used the parameters
$C = 0.7$ and $\alpha = 0.85$ (Line~\ref{line:main2progressiveWidening} of Alg.~\ref{algo:main2}).

The choice of parameters for our two MCTS-based algorithms is as follows:
for each combination with the hill-climbing optimization solvers, we present a set of
parameters that give good results over all the specifications.
This is justified, because the performance is quite dependent on these
parameters, and one choice that works for a given combination of a falsification algorithm and a hill-climbing solver
might just not work for another combination. However, note that we do \emph{not}
change the settings across  the specifications.

As we discuss at the end of~\S{}\ref{subsec:algoMain1}, different timeouts
 are set for hill-climbing in playout (Line~\ref{line:main1playOutHillClimb} of Alg.~\ref{algo:main1}) and to hill-climbing at the end (Line~\ref{line:main1SecondHillClimb} of Alg.~\ref{algo:main1}). Specifically, the timeout for the former is  $\text{TO}_\text{po}$ in Table~\ref{tab:results} (5--15 seconds) while the timeout for the latter is globally 300 seconds.

The results in Table~\ref{tab:results} indicate, at a high-level, that for seemingly
hard problems, the benefit of the extra exploration done by the MCTS layer
significantly increases the falsification rate. This is most evident in S4 and S5, where Breach (with any of CMA-ES, GNM or SA) has at most 30--40\% success rates. Our MCTS enhancements succeed much more often.

For easy problems, the increased exploration typically increases the falsification
times, which is expected.  One reason is that falsification is in general a hard problem that can only be tackled by heuristics. 
We note from Table~\ref{tab:results} that the additional execution time is often not prohibitively large.
We also note that  there is generally no single algorithm that works on all instances equally well.
For example, for Sstable, both Breach and our algorithms are even weaker than random testing. However, our algorithms still increase the falsification rate compared to Breach.

The choice of a hill-climbing optimization solver has a great influence on the outcome.
CMA-ES has built-in support for some exploration before the search converges in
the most promising direction. Nevertheless, we see that the upper-layer optimization by MCTS can improve success rates (S4, S5, Sstable). 
The Nelder-Mead variant GNM has very little
support for exploration and furthermore, Breach's implementation is not stochastic
(it uses deterministic low-discrepancy sequences as a source of quasi-randomness).
For this reason, the method quickly converges to non-falsifying minima that are local
and cannot be escaped without extra measures. Thus, using MCTS  pays off especially with GNM; see for
example  S3 and S4. Conversely, SA heavily relies on exploration and keeps
just a single good trace found so far, limiting its exploitation.
In combination with MCTS, SA shows mixed performance. In some cases
falsification time becomes longer (S1, S3), whereas for S4, MCTS is able to overcome
this particular limitation, presumably because it maintains several good prefixes.
For the free floating robot, we observe that our approach needs additional time
in comparison to Breach with CMA-ES (within an order of magnitude),
which is reasonable given the added exploration
on the exponentially larger state space. However, it does increase the falsification rate with GNM and SA, for the same reasons
as before.

The difference between the two variants, Algs.~\ref{algo:main1}
and~\ref{algo:main2} (the latter with progressive widening),
is not significant on most of the examples. However, progressive widening has a
positive effect on the success rate and falsification time for~S2 and ~S5.

In the experiments, we set the MCTS budget (number of iterations of the main loop) to be 20--40. Note that the number of all possible nodes is much greater: it is
 $(1 + |A| + |A|^2 + \cdots + |A|^K)$. For AT and AFC (2 input signals, $L = 3\times 5$ and $K = 5$), it is $813616$; and for FFR (4 input signals, $L = 2\times2\times2\times2$ and $K = 3$), it is 4369. The overall success rates seem to suggest that, not only in computer Go but also in hybrid system falsification, MCTS is very effective in searching in a vast space with  limited resources.

\subsection{Evaluation of Parameter Choices}
\label{sec:parameters}

We evaluate the effect of the parameters using the specification S4 for the
AT model, where the success of falsification varies strongly.
For the experiments in this section we focus on Alg.~\ref{algo:main1} (Basic).

Table~\ref{tab:results:scaler} contains 4 sub-tables, each showing the results for the different optimization
solvers when varying a hyperparameter.

The first concern is about the scalar $c$ for exploration/exploitation.
We observe that there is a general trend that falsification rate improves with
increased focus on exploration.
It is particularly evident when comparing the results of $c = 0.02$ and $c = 0.5$.
However,  no significant performance gap is observed between $c = 0.5$ and $c = 1.0$,
indicating that $c = 0.5$ is already sufficient for optimization solvers to benefit from exploration.

Next, consider the results for different partitioning of the input space,
where $L = n \times m$ means that the throttle range is partitioned
into $n$~actions and the brake range into $m$~actions
(for the AT model; pedal and engine for the AFC model).
We note that the different choices have much less influence than the scalar~$c$. However, there are some differences, for example GNM seems to cope badly with
the coarse partitioning $2 \times 2$ in the first column, which could be
attributed to its reliance on guidance by the MCTS layer.

With respect to the timeout for individual playouts TO$_\text{PO}$,
we observe that it is correlated with overall falsification time.
This is expected, as we spend more time in non-falsifying regions of the input
space as well.

Varying the number of control points $K$ (and therefore the depth of the MCTS tree),
shows that for the respective requirement, $K = 3$ is insufficient but the
results for more control points are not clear. As more control points make the
problem harder due to the larger search space, the falsification rate drops
(specifically for $K = 10$). Note that we purposely keep the MCTS budget and
playout time consistent to expose this effect, whereas in practice one might
want to increase the limits when the problem is more complex.

\begin{table}[t]
\setlength\tabcolsep{3pt}
\caption{Parameter Variation for Alg.~\ref{algo:main1} (Basic)
         \normalfont
 Success rate and average time (in seconds, only successful trials)
for 4 parameter variations, respectively scalar $c$, input space partition $L$, 
playout timeout $\text{TO}_{\text{po}}$ and the number $K$ of control points. 
The default parameter settings are: maximum tree size (MCTS budget) is 60, and $c = 0.2$,  $L = 2\time2$, $\text{TO}_{\text{po}} = 10$, $K = 5$ (gray headed columns). The green backgrounded cells are the best performers w.r.t.
each solver.}
\label{tab:results:scaler}
\small
\begin{tabular}{lrrrrrrrrrr}
\toprule
    & \multicolumn{2}{c}{$c = 0.02$}
    & \multicolumn{2}{c}{\tbgray$c = 0.2$}
    & \multicolumn{2}{c}{$c = 0.5$}
    & \multicolumn{2}{c}{$c = 1.0$} \\
      Solver
    & succ. & time
    & succ. & time
    & succ. & time
    & succ. & time \\
\cmidrule(r){1-1}
\cmidrule(lr){2-3}
\cmidrule(lr){4-5}
\cmidrule(lr){6-7}
\cmidrule(lr){8-9}
CMA-ES & 6/10 & 826.1 &   7/10 &  728.7 & 8/10 & 725.7 & \tbcolor 9/10 & \tbcolor 744.3 \\
GNM    & 0/10 &      - &   \tbcolor4/10 & \tbcolor 807.3 & 3/10 & 779.4 & 3/10 & 791.4 \\
SA     & 1/10 & 719.5 & 8/10 & 733.5 & \tbcolor 9/10 & \tbcolor 736.3 & 8/10 &  799.1 \\ \hline\hline

     & \multicolumn{2}{c}{\tbgray$L = 2\times 2$}
    & \multicolumn{2}{c}{$L = 3\times 3$}
    & \multicolumn{2}{c}{$L = 3\times 5$}
    & \multicolumn{2}{c}{$L = 5\times 5$} \\
     Solver
     & succ. & time
    & succ. & time
    & succ. & time
    & succ. & time \\
\cmidrule(r){1-1}
\cmidrule(lr){2-3}
\cmidrule(lr){4-5}
\cmidrule(lr){6-7}
\cmidrule(lr){8-9}
CMA-ES&7/10 & 728.7 & \tbcolor 9/10 &\tbcolor 674.4 &  9/10 & 740.2 & 8/10 & 743.4 \\
GNM&    4/10 & 807.3 & 3/10 & 712.3 & 9/10 & 721.6 & \tbcolor10/10 & \tbcolor724.2 \\
SA&       \tbcolor 8/10 & \tbcolor733.5 & 6/10 & 755.7 &  8/10 & 832.0 & 6/10 & 832.8 \\ \hline\hline
& \multicolumn{2}{c}{$\text{TO}_{\text{po}} = 5$}
    & \multicolumn{2}{c}{\tbgray$\text{TO}_{\text{po}} = 10$}
    & \multicolumn{2}{c}{$\text{TO}_{\text{po}} = 15$}
    & \multicolumn{2}{c}{$\text{TO}_{\text{po}} = 20$} \\
     Solver
     & succ. & time
    & succ. & time
    & succ. & time
    & succ. & time \\
\cmidrule(r){1-1}
\cmidrule(lr){2-3}
\cmidrule(lr){4-5}
\cmidrule(lr){6-7}
\cmidrule(lr){8-9}
CMA-ES&8/10 & 431.8 & 7/10 & 728.7 &  \tbcolor9/10 & \tbcolor776.2 & 7/10 & 1330.1 \\
GNM&    3/10 & 502.6 & \tbcolor4/10 & \tbcolor807.3 & 4/10 & 809.4 & 2/10 & 1397.1 \\
SA&        7/10 & 510.5 & \tbcolor8/10 & \tbcolor733.5 &  7/10 & 1108.0 & 8/10 & 1342.5 \\ \hline\hline

    & \multicolumn{2}{c}{$K = 3$}
    & \multicolumn{2}{c}{\tbgray$K = 5$}
    & \multicolumn{2}{c}{$K = 7$}
    & \multicolumn{2}{c}{$K = 10$} \\
     Solver
     & succ. & time
    & succ. & time
    & succ. & time
    & succ. & time \\
\cmidrule(r){1-1}
\cmidrule(lr){2-3}
\cmidrule(lr){4-5}
\cmidrule(lr){6-7}
\cmidrule(lr){8-9}
CMA-ES&0/10 &   -  & \tbcolor7/10 &\tbcolor 728.7 &  6/10 & 711.5 & 5/10 & 777.9 \\
GNM    &0/10 &     - & 4/10 & 807.3  &  1/10 & 664.3 & \tbcolor6/10 & \tbcolor892.8 \\
SA       &0/10 &      - & 8/10 &  733.5   &  \tbcolor 8/10 &\tbcolor 709.7 & 3/10 & 750.9 \\
\bottomrule
\end{tabular}
\end{table}

\section{Related Work}\label{sec:relatedwork}

Formal verification approaches to correctness of hybrid systems employ a wide range of techniques, including model checking, theorem proving, rigorous numerics, nonstandard analysis, and so on~\cite{ChenAS13,GaoAC12,FrehseGDCRLRGDM11,FanQM0D16,DreossiDP16,Platzer10Book,HasuoS12CAV}.
These are currently not very successful in dealing with complex real-world systems, due to issues like  scalability and black-box components. 

Optimization-based falsification of hybrid systems has therefore  attracted attention as a testing technique that adaptively searches for  error input using algorithms, using recent advances in machine learning. An overview is given in~\cite{KapinskiDJIB16}.

We now discuss the relationship between the current work and existing works in the context of falsification.

Monte Carlo sampling is used in~\cite{NghiemSFIGP10} for falsification. Our thesis is that Monte Carlo \emph{tree} search---an extension of Monte Carlo methods---yields a powerful guiding method in optimization-based falsification.

The so-called \emph{multiple-shooting} approach to falsification is studied in~\cite{ZutshiDSK14}. It consists of: an upper layer that searches for an \emph{abstract error trace} given by a succession of cells; and a lower layer, where an abstract error trace is concretized to an actual error trace by picking points from cells. This two-layered framework differs from ours:
they focus on safety specifications (avoiding an unsafe set); this restriction allows  search heuristics 
that rely on spacial metrics (such as  $\mathrm{A^{*}}$ search). In our current work, we allow arbitrary $\STL$ specifications and use robustness values as guidance. Our framework can be seen as an integration of multiple-shooting (the upper layer) and single-shooting (the lower layer); they are interleaved in the same way as search and sampling are interleaved in MCTS.

Besides MCTS, \emph{Gaussian process learning (GP learning)} has also attracted attention in machine learning as a clean way of balancing exploitation and exploration. The GP-UCB algorithm is a widely used strategy there. Its use in hybrid system falsification is pursued e.g.\ in~\cite{AkazakiKH17,SilvettiPB17}.

The value of exploration/coverage has been recognized in the falsification community~\cite{KuratkoR14,DeshmukhJKM15,AdimoolamDDKJ17,DreossiDDKJD15}, not only for efficient search for error inputs, but also for correctness guarantees in case no error input is found. In this line, the closest to the current work is~\cite{AdimoolamDDKJ17}, in which search is guided by a coverage metric  on input spaces. 
 The biggest difference in the current work is that we structure the input space by time, using time stages (see Fig.~\ref{fig:timeStaging}). We explore this staged input space in the disciplined manner of MCTS.
In~\cite{AdimoolamDDKJ17} there is no such staged structure in input spaces, and they use support vector machines (SVM) for identifying promising regions.
Underminer~\cite{BalkanTDJK18} is a falsification tool that learns the
(non-)convergence of a system to direct falsification and parameter mining.
It supports STL formulas, SVMs, neural nets, and Lyapunov-like functions as
classifiers.

Tree-based search is also used in~\cite{DreossiDDKJD15} for falsification.  They use \emph{rapidly-exploring random trees} (RRT), a technique widely used for path planning in robotics. Their use of trees is geared largely towards exploration, using the coverage metric called \emph{star discrepancy} as guidance. In their algorithm, robustness-guided hill-climbing optimization plays a supplementary role.
This is in contrast to our current framework, where we use MCTS and  systematically integrate it with  hill-climbing optimization.

Many works in coverage-guided falsification~\cite{KuratkoR14,DreossiDDKJD15} use metrics in the space of output or internal states, instead of the input space. A challenge in such methods is that, in a complex model, the correlation between input and output/state is  hard to predict. It is hard to steer the system's output/state to a desired region.

There have been efforts to enhance expressiveness of MTL and STL, so that engineers can express richer intentions---such as time robustness and frequency---in specifications~\cite{AkazakiH15,NguyenKJDBJ17}. This research direction  is orthogonal to ours; we are able to investigate the use of such logics in our current framework. Other recent works with which our current results could be combined include~\cite{HoxhaDF18}, which mines parameter regions, and~\cite{DreossiDS17} that aims to exploit features of machine learning components of system models for the sake of falsification.

We believe that the combination of MCTS 
and application-specific lower-layer optimization---an instance of which is the proposed falsification framework---is a general methodology
applicable to a variety of applications. For example, for the MaxSAT
problem, the work~\cite{GoffinetR16} uses MCTS combined with hill-climbing local optimization.

Use of MCTS for search-based testing of hybrid systems is pursued
in~\cite{lee2015adaptive}.
We differ  from~\cite{lee2015adaptive} in the target systems: ours are
deterministic, while~\cite{lee2015adaptive} searches for random seeds
for stochastic systems. We also  combine robustness-guided hill-climbing
optimization.

There are strong similarities between our falsification approach using MTCS and statistical model checking (SMC) using \emph{importance splitting}~\cite{JegourelLegaySedwards2013}.
The robustness semantics of STL can be seen as a ``heuristic score function''~\cite{JegourelLegaySedwards2014},
with both approaches using time staging and the notion of ``levels'' to iteratively guide the search to more promising subspaces.
The principal difference is that importance splitting randomly explores a diverse set of traces that satisfy the property (in order to reduce the variance of the estimate of its probability), while our falsification approach finds a single falsifying input by optimizing (exploiting) the results of random exploration.

\section{Conclusions and Future Work}\label{sec:concl}
In this work we have presented a two-layered optimization framework for hybrid system falsification. It combines Monte Carlo tree search---a widely used stochastic search method that effectively balances exploration and exploitation---and hill-climbing optimization---a  local search method whose use in hybrid system falsification is established in the community. Our experiments demonstrate its promising performance.

In addition to the future work already outlined in~\S{}\ref{sec:relatedwork}, we add the following.

We have shown how systematic exploration can improve the chances of finding error input.
Such exploration can also be used as a measure of confidence about a system's validity, in the case that no error input is found (see~\cite{AdimoolamDDKJ17,DreossiDDKJD15} and~\S{}\ref{sec:relatedwork}).
Concretely, it would be interesting to compute a quantitative coverage metric from the result of our MCTS algorithm.

Our choice of simple grid partitioning of actions in MCTS search trees achieves good performance.
Other choices are also possible, such as using the extension of MCTS to continuous action sets in~\cite{Coulom07,CouetouxHSTB11}.

Finally, an extension of our framework to stochastic hybrid systems does not seem hard, following the
MCTS approach in~\cite{lee2015adaptive} that uses models with direct
access to randomness seeds (see~\S{}\ref{sec:relatedwork}).

\section*{Acknowledgment}
Thanks are due to Georgios Fainekos,  Fuyuki Ishikawa,  Masaaki Konishi  and Akihisa Yamada for helpful discussions. 
The authors are supported by ERATO HASUO Metamathematics for Systems Design Project
(No.~{JPMJER1603}),
JST;
and Grants-in-Aid {No.~15KT0012},
{JSPS}.

\ifCLASSOPTIONcaptionsoff
  \newpage
\fi

\bibliographystyle{IEEEtran}
\bibliography{references}

\end{document}